\documentstyle[11pt,epsf]{article}

\def\kms{\hbox{km s$^{-1}$}}
\def\VLSR{\hbox{$V_{\rm LSR}$}}

\def\sun{\hbox{$\odot$}}

\def\lesssim{\mathrel{\hbox{\rlap{\hbox{\lower4pt\hbox{$\sim$}}}\hbox{$<$}}}}
\def\gtrsim{\mathrel{\hbox{\rlap{\hbox{\lower4pt\hbox{$\sim$}}}\hbox{$>$}}}}

\def\arcdeg{\hbox{$^\circ$}}

\def\arcsec{\hbox{$^{\prime\prime}$}}

\newlength{\minitwocolumn}
\begin{document}
\pagestyle{empty}

\begin{flushleft}

\vspace*{2cm}

{\Large\bf A LARGE-SCALE CO IMAGING OF THE GALACTIC CENTER \par
II. DYNAMICAL PROPERTIES OF MOLECULAR CLOUDS}

\vspace{1cm}

{T. Oka$^1$, T. Hasegawa$^2$, F. Sato$^3$, M. Tsuboi$^4$, and A. Miyazaki$^4$}\\

\begin{description}
\labelsep=0ex
\itemsep=-0.7ex
\item[$^1$]{\it The Institute of Physical and Chemical Research (RIKEN), 
2-1 Hirosawa, Wako, Saitama 351-0198, Japan}\\
\item[$^2$]{\it Institute of Astronomy, Faculty of Science, 
The University of Tokyo, 2-21-1 Osawa, Mitaka, Tokyo 181-8588, Japan}\\
\item[$^3$]{\it Department of Astronomy and Earth Sciences, 
Tokyo Gakugei University, 4-1-1 Nukui-kita, Koganei, Tokyo 184-8501, Japan}\\
\item[$^4$]{\it Institute of Astrophysics and Planetary Science, 
Ibaraki University, 2-1-1 Bunkyo, Mito, Ibaraki 310-8512, Japan}\\
\end{description}

\end{flushleft}

\vspace{1cm}

\parskip=1ex
\baselineskip=2.5ex

{\large ABSTRACT}\par

The data from the Nobeyama Radio Observatory 45 m telescope Galactic Center CO survey  
have been analyzed to generate a compilation of molecular clouds with intense CO emission in this region.  
Clouds are identified in an automated manner throughout the main part of the survey data 
for all CO emission peaks exceeding 10 K ($T_R^*$).  

Correlations between the size, velocity dispersion, virial mass, and the CO luminosity, 
for the molecular clouds in the Galactic center were shown.   
We diagnosed gravitational stabilities of identified clouds 
assuming that the disk clouds are nearly at the onset of gravitational instability.  
Most of the clouds and cloud complexes in the Galactic center are gravitationally stable, 
while some clouds with intense CO emission are gravitationally unstable.  

\vspace{2ex}

{\large INTRODUCTION}\par

The structure and statistical properties of molecular clouds could be important clues 
to understanding the mechanisms of formation and evolution of molecular clouds, and formation of stars within them.  
The identification of molecular clouds in the Galactic disk has shown their size and mass spectra, 
as well as correlations between the size, velocity dispersion, virial mass, and the CO luminosity, 
(Scoville et al. 1987; Solomon et al. 1987; Dame et al. 1986).  

Molecular clouds in the central 500 pc of the Galaxy are characterized by large velocity widths, 
high temperaure, and high density, which differ considerably from those in the Galactic disk 
(see e.g., G\"usten 1989, and references therein). 
It must be meaningful to compare the gross properties of molecular clouds 
in this highly turbulent region with those in the Galactic disk.  

The recently completed large-scale, high resolution CO {\it J}=1--0 survey of the Galactic center 
with the NRO 45 m telescope provides extensive data sets 
which are ideal for studies of cloud properties (Oka et al. 1998b).  
Here we present preliminary results of the analyses based on cloud identification 
and discuss their gravitational stabilities.  
The distance to the Galactic center was assumed to be $D\!=\!8.5$ kpc.  

\vspace{2ex}

{\large CLOUD DEFINITION}\par

A $(l, b, V)$ 3-dimensional CO data cube which covers the region 
$-0.8\arcdeg\!\leq\!l\!\leq\!+1.7\arcdeg$, $-0.38\arcdeg\!\leq\!b\!\leq\!+0.38\arcdeg, 
|\VLSR|\!\leq\!200$ \kms\ was used for the cloud search.  
The data were smoothed to a 60\arcsec\ spatial and a 2 \kms\ velocity resolution (FWHM), 
and resampled onto a $34\arcsec\!\times\!34\arcsec\!\times\!2$ \kms\ grid.  

\begin{minipage}[t]{8.8cm}
\parskip=1ex
\baselineskip=2.5ex

Clouds are defined as topologically closed surfaces of antenna temperature in $(l, b, V)$ space.  
We took three boundary threshold intensities $T_{\rm min}\!=\!5,7.5,10$ K, 
and required minimum peak temperatures twice the thresholds,  
$T_{\rm peak}\!\geq\!10,15,20$ K, respectively.  

We identified 165 clouds in the 45 m data set.  
Although some clouds appear repeatedly at different thresholds, 
we leave them as separate entries to see cloud properties of various scales. 
Clouds with peak velocities between $-60$ and  $+20$ \kms\ are excluded from the analyses 
to avoid the severe contamination of molecular gas in the Galactic disk.  
The same procedure has been applied also for the more extensive CO data obtained 
by the 1.2 m Southern mm-wave telescope at CTIO (Bitran et al. 1997), 
with $T_{\rm min}\!=\!1, 2.5$ K and $T_{\rm peak}\!\geq\!2, 5$ K, respectively.  
47 clouds are identified in the 1.2 m data set.  

\vspace{2ex}

{\large CLOUD PROPERTIES}\par

Figure 1 shows a size ($S\!\equiv \! D {\rm tan}\sqrt{\sigma_l \sigma_b }$) 
versus velocity dispersion ($\sigma_V$) plot for the GC clouds 
with the same plot for the disk clouds.   
Almost all the GC clouds identified have larger velocity widths 
above the $S$-$\sigma_V$ line of the disk clouds.  
A linear least-squares regression in the log leads to the result 
\begin{equation}	
\sigma_V = 2.7^{+0.4}_{-0.3}\; S^{0.48\pm 0.07}\;\; (\kms ),  
\end{equation}  
for the 45 m GC clouds.  

Figure 2 shows a plot of CO total luminosity 
($L_{\rm CO}\!\equiv\!D^2 \int\!\!\!\int\!\!\!\int T_R^{*} d\Omega dV$) 
versus virial theorem mass ($M_{\rm VT}\!\equiv\!3 f_p {S \sigma_V^2}/{G}$) 
for the GC clouds with the same plot for the disk clouds. 
Again, most of the GC clouds identified have larger virial theorem masses  
above the $L_{\rm CO}$-$M_{\rm VT}$ line of the disk clouds.  
A linear least-squares regression leads to the result 
\begin{equation}	
M_{\rm VT} = 56^{+29}_{-19}\; (L_{\rm CO})^{0.85\pm 0.04}\;\; (M_{\sun}),   
\end{equation}
for the 45 m GC clouds.  
$M_{\rm VT}$-$L_{\rm CO}$ loci of less luminous GC clouds are similar to the disk clouds, 
while they shift upward with increasing $L_{\rm CO}$ and with increasing dispersion in $M_{\rm VT}$.  

\vspace{1cm}

\end{minipage}%

\hspace{9.2cm}%
\begin{minipage}[t]{8.8cm}
\vspace{-20cm}
\vspace{7.5cm}
\includegraphics{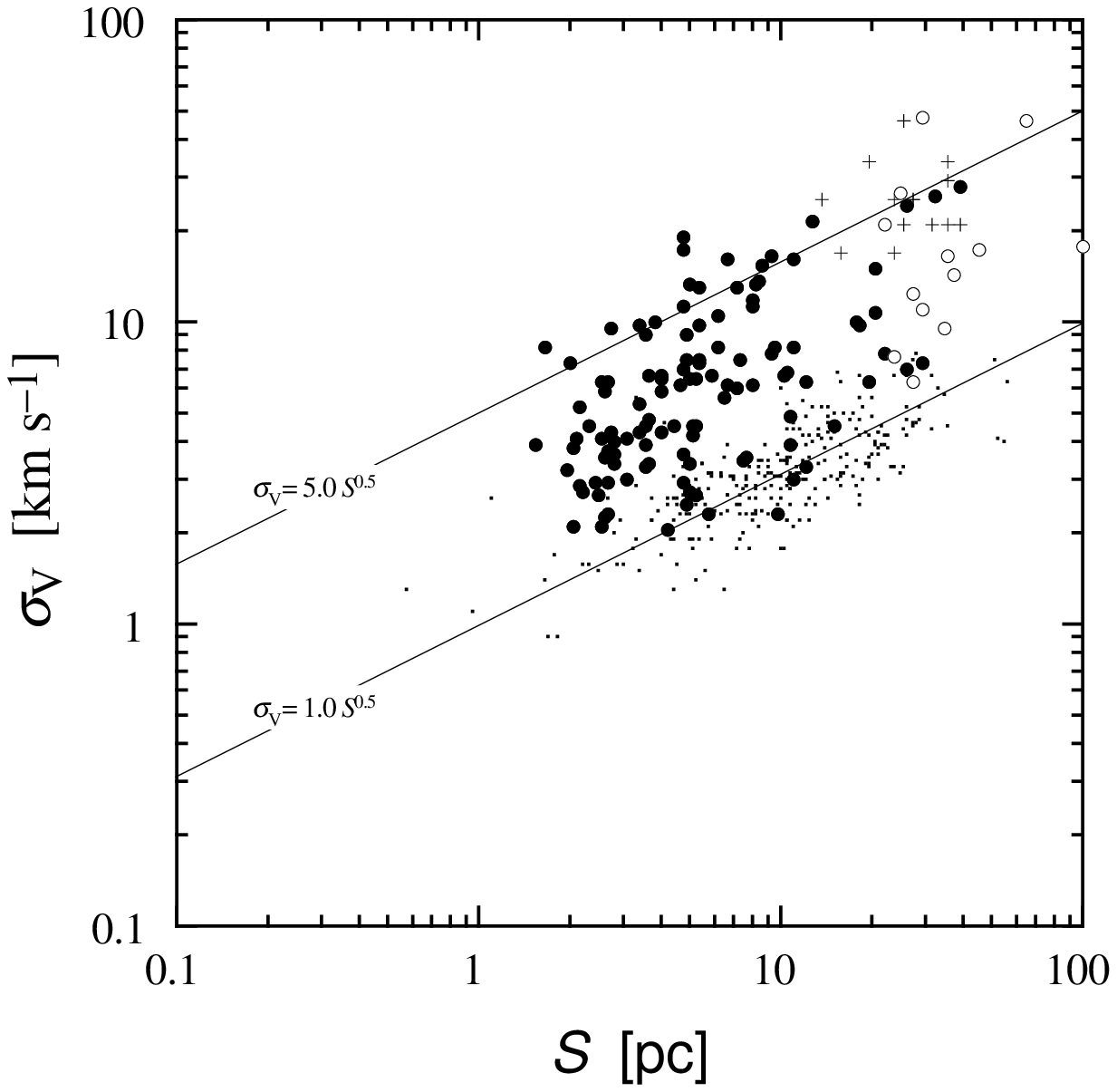}
\footnotesize 
Fig.1 --- $S$-$\sigma_V$ plot is shown for the clouds identified 
both in the 45 m (filled circles) and 1.2 m (open circles) data sets, 
with the same plots for the disk clouds (dots, Solomon et al. 1987) 
and the large GC clouds identified manually in the CO {\it J}=2--1 survey data (crosses, Oka et al. 1998a).  
Best fit lines to the data sets of Solomon et al. (1987) and of Oka et al. (1998a) are also shown. 

\vspace{8.5cm}
\includegraphics{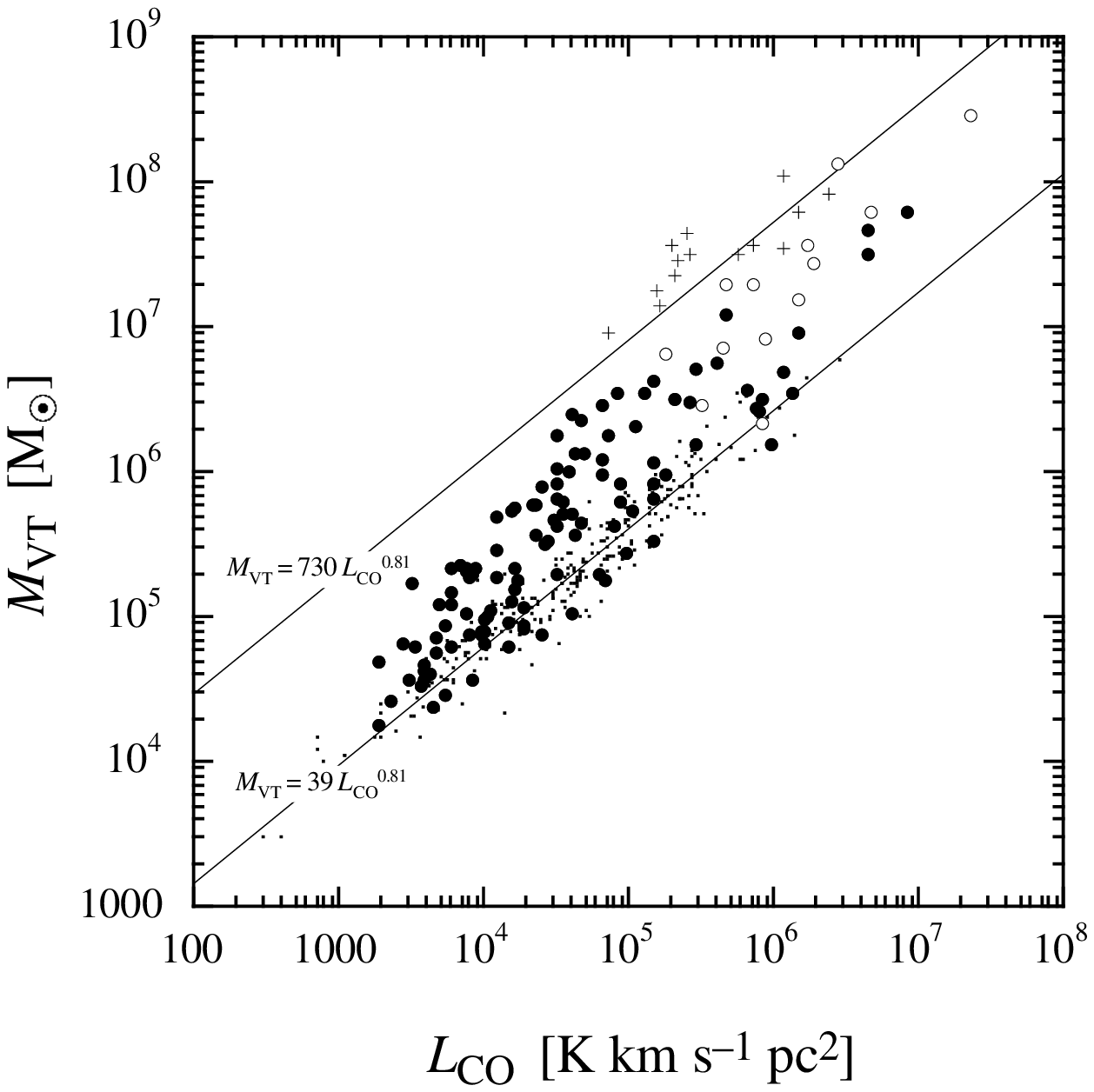}
\footnotesize 
Fig.2 --- $L_{\rm CO}$-$M_{\rm VT}$ plot is shown for the GC clouds and the disk clouds.  
Marks are the same as Fig.1.  
Best fit lines derived by Solomon et al. (1987) and by Oka et al. (1998a) are also shown.
\end{minipage}

\vspace{2ex}

{\large CLOUD IN EQUILIBRIUM WITH EXTERNAL PRESSURE}\par

The larger $M_{\rm VT}$ values of the GC clouds than those of the disk clouds with the same $L_{\rm CO}$ 
can not be solved by introducing a large CO-to-H$_2$ conversion factor, 
since $\gamma$-ray (Blitz et al. 1985), far-infrared (Sodroski et al. 1995), 
and X-ray observations (Sakano et al. 1997) 
all suggest a CO-to-H$_2$ conversion factor in the Galactic center smaller than the standard value.  

The virial equation for a spherically symmetric non-magnetic cloud in an equilibrium state. 
\begin{equation}
3 \sigma_V^2 M - a \frac{G M^2}{R} = 4 \pi R^3 p
\end{equation}
gives two equilibrium masses, 
\begin{eqnarray}
M_{\rm VT} & = & \alpha M_0, \\
\alpha & = & \alpha_{\pm} \equiv (1 \pm \sqrt{1-\beta})/2 
\end{eqnarray}

\begin{minipage}[t]{8.8cm}
\parskip=1ex
\baselineskip=2.5ex

where $\beta\!=\!16\pi a R^2 p/9 \sigma_V^4$, 
and $M_0\!=\!3R\sigma_V^2/G$ is the commonly used virial theorem mass.  
For an isothermal cloud, an equilibrium state with $\alpha\!=\!\alpha_+\,(>\!1/2)$ is gravitationally unstable  
and with $\alpha\!=\!\alpha_-\,(<\!1/2)$ is gravitationally stable.

For an opaque cloud with uniform CO brightness of $T_{\rm CO}$, total CO luminosity can be expressed as 
$L_{\rm CO}\!=\!2 \pi R^2 T_{\rm CO} \sigma_V$.  
From this expression, and equating the virial mass $M_{\rm VT}\!=\!\alpha M_0$ to $(4/3)\pi R^3 \rho$, 
where $\rho$ is the mean mass density, 
we get the linear $L_{\rm CO}$-$M_{\rm VT}$ relation, 
\begin{equation}
M_0 = \alpha^{-\frac{1}{2}} \left[\left(\frac{\rho}{\pi a G}\right)^\frac{1}{2} \frac{1}{T_{\rm CO}}\right] L_{\rm CO} .
\end{equation}
On the other hand, the total molecular mass including the helium correction 1.36 is written as  
\begin{equation}
M = 2.2\times 10^{-20} X L_{\rm CO}\;\; (M_{\sun})
\end{equation} 
where $L_{\rm CO}$ is in K \kms\ pc$^2$  
and introducing the CO-to-H$_2$ conversion factor $X\!\equiv\!N({\rm H}_2)/I_{\rm CO}$.  
Using the virial mass $M_{\rm VT}\!=\!\alpha M_0$ as the total molecular mass, 
and comparing eqs.(6) and (7), we get 
\begin{equation}
M_0 = 2.2\times 10^{-20} \alpha^{-\frac{1}{2}} X_0 L_{\rm CO} .  
\end{equation}
The large variation in $L_{\rm CO}$-$M_{\rm VT}$ relation is explained 
by choosing different $\alpha$ in the eq.(8) for each region or scale (Oka et al. 1998a).  

\vspace{1cm}

\end{minipage}

\hspace{9.2cm}%
\begin{minipage}[t]{8.8cm}
\vspace{-15cm}
\vspace{13.5cm}
\includegraphics{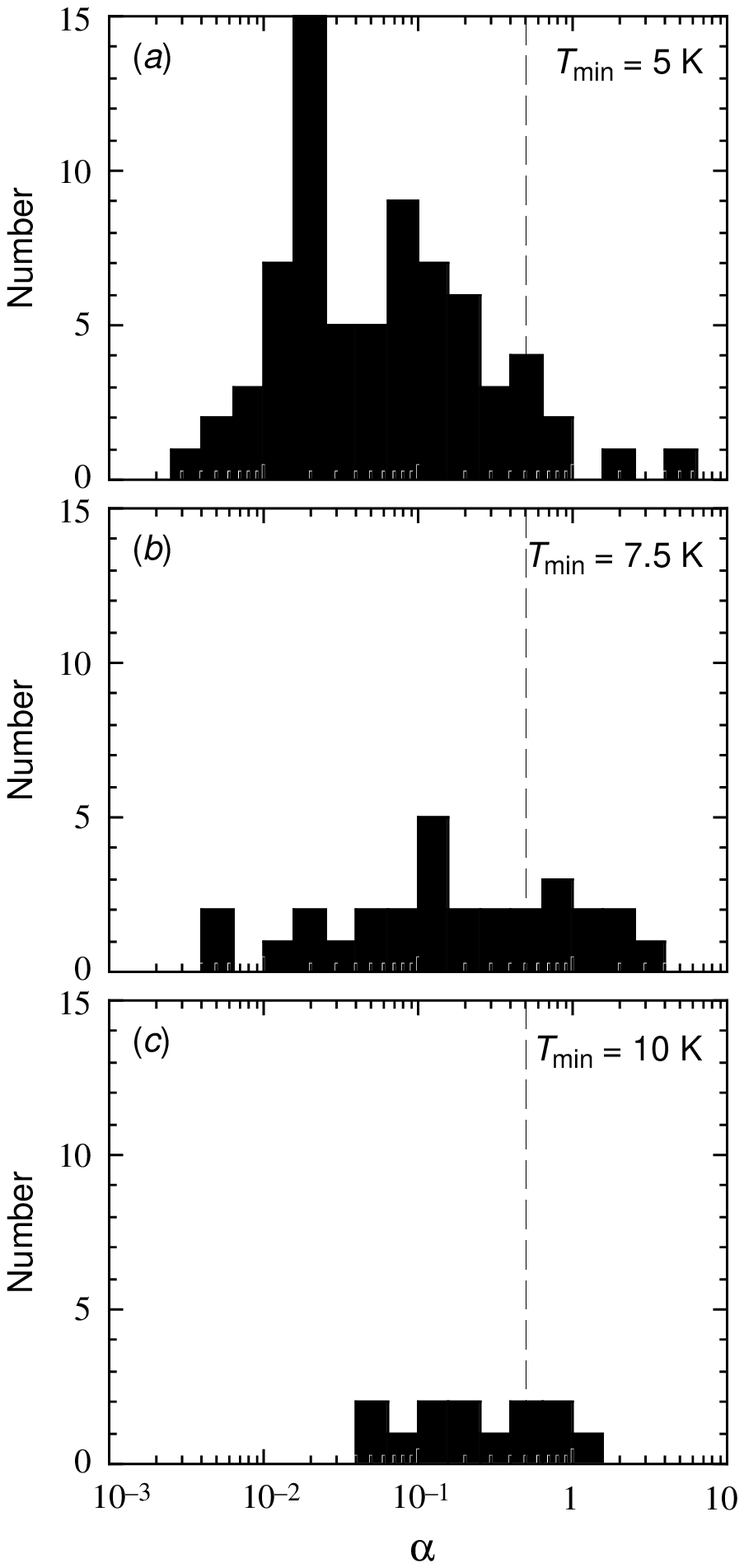}
\footnotesize 
Fig.3 --- Distributions of $\alpha$ are shown for the 45 m GC clouds identified at boundary intensities 
({\it a}) 5 K, ({\it b}) 7.5 K, and ({\it c}) 10 K, respectively.  
Broken lines show the critical state for gravitational instability, $\alpha\!=\!1/2$.  
\end{minipage}

\vspace{2ex}

{\large GRAVITATIONAL STABILITIES OF MOLECULAR CLOUDS}\par

Assuming that the disk clouds are nearly at the onset of gravitational instability (e.g., Chi\`eze 1987), 
$\alpha\!=\!1/2\,(\beta\!=\!1)$, we get $X_0\!=\!1.6\times 10^{20}$ $\mbox{cm}^{-2}\,(\mbox{K}\,\kms )^{-1}$,  
and $X\!=\!1.2\times 10^{20}$ $\mbox{cm}^{-2}\,(\mbox{K}\,\kms )^{-1}$ for the disk clouds,
which is similar to that derived for the Orion molecular clouds from $\gamma$-ray observations (Digel, Hunter, \&\ Mukherjee 1995).  

Gravitational stability of each GC cloud can be diagnosed using the above $X_0$.  
Fig.3 shows the distributions of $\alpha$ for the 45 m GC clouds identified at three boundary thresholds.  
The distribution of $\alpha$ shifts upward with increasing $T_{\rm min}$.  
Clouds identified at $T_{\rm min}\!=\!7.5, 10$ K contain a larger percentage of 
gravitationally unstable ($\alpha\!>\!1/2$) clouds than those at $T_{\rm min}\!=\!5$ K.  
This means that clouds with intense CO emission tend to be gravitationally unstable, 
while most clouds with less intense emission and the complexes of clouds are gravitationally stable.  

Fig.4 shows the {\it l-V} distributions of the GC 45 m clouds with three $\alpha$ ranges.  
It is evident that gravitationally unstable ($\alpha\!>\!1/2$) clouds follow the main ridge of intense CO emission, 
part of which defines two rigidly-rotating molecular arms (Sofue 1995).  
Clouds with $0.05\!\leq\!\alpha\!<\!0.5$ exhibit a distribution similar to gravitationally unstable clouds, 
favoring the main ridge of CO emission with a slight dispersion.  
On the other hand, clouds with $\alpha\!<\!0.05$ spread over the {\it l-V} plane.  
These results suggest that mechanisms such as orbit crowding at the inner Lindblad resonance may play a role to promote 
gravitational instability of clouds and subsequent star formation.  

\vspace*{6cm}
\includegraphics{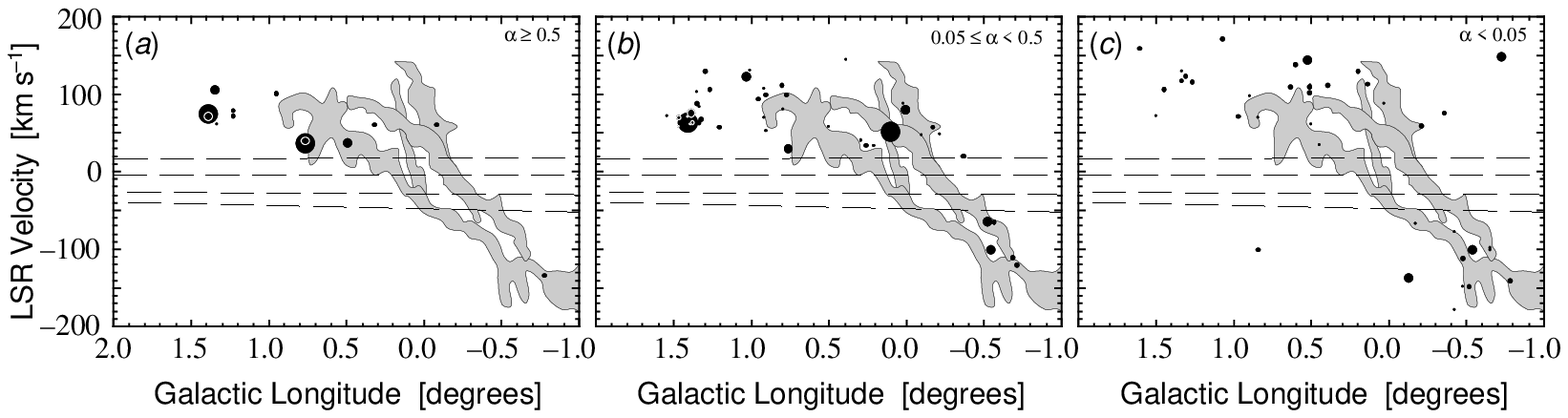}
{\footnotesize 
Fig.4 --- The {\it l-V} distribution of the 45 m GC clouds with ({\it a}) $\alpha\!\geq\!0.5$, 
({\it b}) $0.05\!\leq\!\alpha\!<\!0.5$, and ({\it c}) $\alpha\!<\!0.05$.  
Sizes of circles denote their CO luminosities, 
$L_{\rm CO}\!=\!10^{3-4},\!10^{4-5},\!10^{5-6},\!10^{6-7}$ K \kms\ pc$^2$, with increasing size, respectively. 
Shaded areas denote Sofue's molecular arm I--IV.  
Broken lines show the {\it l-V} loci of the Galactic arms.  
}

\vspace{2ex}

{\large CONCLUSIONS}\par

The major results of our analyses are the following:

1. The velocity width of clouds in the Galactic center region correlates with their size 
in the same way as those of the disk clouds, $S\!\propto\!\sigma_V^{0.5}$.  
The size$-$line-width coefficient ($\sigma_V/S^{0.5}$) of the GC clouds 
is about 2.5 times larger than that of the disk clouds.  

2. The CO luminosity-virial theorem mass law for the GC clouds is $M_{\rm VT}\!\propto\!(L_{\rm CO})^{0.85}$, 
which is also similar to that of the disk clouds.  
Most of the GC clouds identified have larger virial theorem masses  
above the $L_{\rm CO}$-$M_{\rm VT}$ line of the disk clouds.  
$M_{\rm VT}$-$L_{\rm CO}$ loci of less luminous GC clouds are similar to the disk clouds, 
while they shift upward with increasing $L_{\rm CO}$ and with increasing dispersion in $M_{\rm VT}$.  

3. Gravitational stability of the individual GC clouds was diagnosed.  
Most of the GC clouds are gravitationally stable, 
being in equilibrium with the external pressure of intercloud medium, 
while some clouds with intense CO emission are gravitationally unstable.  

4. The gravitationally unstable clouds in the Galactic center follow the main ridge of intense CO
emission,  part of which defines two rigidly-rotating molecular arms (Sofue 1995).  
This suggests that mechanisms such as orbit crowding at the inner Lindblad resonance could promote 
gravitational instability.  

\vspace{2ex}


{\large REFERENCES}\par
\vspace{-2ex}
\begin{description}
\labelsep=0ex
\itemsep=-0.5ex
\item[]Bitran, M., Alvarez, H., Bronfman, L., May, J., \&\ Thaddeus, P., {\it Astron. Astrophys. Suppl.}, {\bf 125}, 99 (1997)  
\item[]Blitz, L., Bloemen, J. B. G. M., Hermsen, W., \&\ Bania, T. M., {\it Astron. Astrophys.}, {\bf 143}, 267 (1985)  
\item[]Chi\`eze, J. P., {\it Astron. Astrophys.}, {\bf 171}, 225 (1987)  
\item[]Dame, T. M., Elmegreen, B. G., Cohen, R. S., \&\ Thaddeus, P., {\it Astrophys. J.}, {\bf 305}, 892 (1986)  
\item[]Digel, S. W., Hunter, S. D., \&\ Mukherjee, R., {\it Astrophys. J.}, {\bf 441}, 270 (1995)
\item[]G\"usten, R., in IAU Symp.136, {\it The Center of the Galaxy}, ed. M. Morris (Dordrecht: Kluwer), 89 (1989)
\item[]Maloney, P., {\it Astrophys. J.}, {\bf 334}, 761 (1988)
\item[]Oka, T., Hasegawa, T., Hayashi, M., Handa, T., \&\ Sakamoto, S.,  {\it Astrophys. J.}, {\bf 493}, 730 (1998a)
\item[]Oka, T., Hasegawa, T., Sato, F., Tsuboi, M., \&\ Miyazaki, A.,  {\it Astrophys. J. Suppl.} in press (1998b) 
\item[]Sakano, M., Nishiuchi, M., Maeda, Y., \&\ Koyama, K., IAU Symp.184, {\it The Central Regions of the Galaxy and Galaxies}, 227 (1997) 
\item[]Scoville, N. Z., Yun, M. S., Clemens, D. P., Sanders, D. B., \&\ Waller, W. H., {\it Astrophys. J. Suppl.}, {\bf 63}, 821 (1987) 
\item[]Sodroski, T. J., Odegard, N., Dwek, E., Hauser, M. G., Franz, B. A., {\it et al.}, {\it Astrophys. J.}, {\bf 452}, 262 (1995) 
\item[]Sofue, Y., {Publ. Ast. Soc. Japan}, {\bf 47}, 527 (1995)
\item[]Solomon, P. M., Rivolo, A. R., Barret, J., \&\ Yahil, A., {\it Astrophys. J.}, {\bf 319}, 730 (1987)
\end{description}

\end{document}